\documentclass[aip,reprint,floatfix,graphicx,superscriptaddress,nofootinbib]{revtex4-1}
\usepackage{graphicx}
\usepackage{color,soul}
\usepackage{amsmath,amssymb}
\usepackage{float}
\usepackage{ulem}

\usepackage[dvipsnames]{xcolor}

\usepackage[utf8]{inputenc}
\usepackage[T1]{fontenc}
\usepackage{mathptmx}

\usepackage{hyperref}
\hypersetup{
    colorlinks=true,
    linkcolor=blue,
    filecolor=magenta,      
    urlcolor=blue,
    citecolor=blue,
}

\usepackage{fancyhdr}
\draft 

\begin{document}






\title{Hybrid InP and SiN integration of an octave-spanning frequency comb} 



\author{Travis C. Briles}
\email{travis.briles@nist.gov}
\affiliation{Time and Frequency Division, Physical Measurement Laboratory, National Institute of Standards and Technology, Boulder, Colorado, 80305, USA}
\affiliation{Department of Physics, University of Colorado, Boulder, Colorado, 80302, USA}

\author{Su-Peng Yu}
\affiliation{Time and Frequency Division, Physical Measurement Laboratory, National Institute of Standards and Technology, Boulder, Colorado, 80305, USA}
\affiliation{Department of Physics, University of Colorado, Boulder, Colorado, 80302, USA}

\author{Lin Chang}
\affiliation{ECE Department, University of California Santa Barbara, Santa Barbara, CA, 93106, USA}

\author{Chao Xiang}
\author{Joel Guo}
\affiliation{ECE Department, University of California Santa Barbara, Santa Barbara, CA, 93106, USA}

\author{David Kinghorn}
\affiliation{ECE Department, University of California Santa Barbara, Santa Barbara, CA, 93106, USA}
\affiliation{Pro Precision Process \& Reliability, PO Box 1285 Goleta, CA 93116, USA}

\author{Gregory Moille}
\author{Kartik Srinivasan}
\affiliation{Microsystems and Nanotechnology Division, Physical Measurement Laboratory, National Institute of Standards and Technology, Gaithersburg, Maryland, 20899-6203, USA}

\author{John E. Bowers}
\affiliation{ECE Department, University of California Santa Barbara, Santa Barbara, CA, 93106, USA}

\author{Scott B. Papp}
\affiliation{Time and Frequency Division, Physical Measurement Laboratory, National Institute of Standards and Technology, Boulder, Colorado, 80305, USA}
\affiliation{Department of Physics, University of Colorado, Boulder, Colorado, 80302, USA}


\date{\today}

\begin{abstract}
Implementing optical-frequency combs with integrated photonics will enable wider use of precision timing signals. Here, we explore the generation of an octave-span, Kerr-microresonator frequency comb, using hybrid integration of an InP distributed-feedback laser and a SiN photonic-integrated circuit. We demonstrate electrically pumped and fiber-packaged prototype systems, enabled by self-injection locking.  This direct integration of a laser and a microresonator circuit without previously used intervening elements, like optical modulators and isolators, necessitates understanding self-injection-locking dynamics with octave-span Kerr solitons. In particular, system architectures must adjust to the strong coupling of microresonator back-scattering and laser-microresonator frequency detuning that we uncover here. Our work illustrates critical considerations towards realizing a self-referenced frequency comb with integrated photonics. 
\end{abstract}
\maketitle 

\section{Introduction 
\label{sec:Introduction}}

Dissipative Kerr solitons are promising optical-frequency-comb sources that would enable numerous real-world applications due to their compatibility with integrated photonics \cite{kippenberg2018dissipative}.  In particular, hybrid integration of the microresonator and a InP pump laser opens the door to compact, low-power soliton-microcomb systems that are driven only by electrical current.  This offers the tantalizing prospect of a deployable microcomb system, which is optimized for $f$--$2f$ stabilization \cite{jones2000carrier} to enable optical-frequency synthesis \cite{spencer2018integrated}, optical-clock metrology \cite{newman2019architecture,drake2019terahertz}, and generation of ultrastable microwave signals using optical frequency division \cite{drake2019terahertz}.  Photonic-chip-based microresonators made from materials like silicon nitride (hereafter SiN) \cite{moss2013new}, tantala \cite{jung2020tantala}, and aluminum gallium arsenide \cite{chang2020ultra,moille2020dissipative}, are compelling as soliton microcomb platforms because they allow direct integration of resonator coupling waveguides and other photonic-circuit elements.  Early self-referencing experiments with chip-based microcombs have used $2f$--$3f$ measurements \cite{brasch2016photonic,brasch2017self} and complicated benchtop equipment.  A more straightforward implementation of soliton self-referencing \cite{briles2018interlocking} involves careful group-velocity dispersion (GVD) design of the microresonator to generate an octave-spanning soliton microcomb. In particular, the possibility exists to leverage dispersive-wave (DW) enhancements, caused by higher than second-order GVD, of the soliton-microcomb spectrum at the critical wavelengths for $f$--$2f$ detection, and to control the carrier-envelope-offset frequency, $f_{\text{ceo}}$ of the soliton microcomb with the lithographic fabrication process of the microresonator. The broadest bandwidth soliton microcombs support dual DW spectra that exceed one octave\cite{li2017stably,pfeiffer2017octave,briles2018interlocking,yu2019tuning}.  

Despite this remarkable progress, the promise of a chip-scale, hybrid-integrated, electrically pumped, octave-bandwidth microcomb has yet to be realized, because of two significant outstanding challenges.  The first challenge arises from the limited optical power available from narrow-linewidth semiconductor lasers.  When combined with the optical losses of traditional laser components, such as optical modulators, isolators, and coupling losses between fiber and chip, insufficient power is available for octave-span microcomb generation. The second challenge arises from limitations in the frequency modulation range and bandwidth of high-power semiconductor lasers that hinders soliton generation in microresonators.  To-date, $f_\text{ceo}$ stabilized microcomb systems\cite{lamb2018optical,briles2017kerr,drake2020thermal} have utilized rapid laser-frequency sweeping to mitigate thermal bistability, which is intrinsic to soliton generation due to the requirement of a modulation-instability (MI) intraresonator precursor. Although, this approach has been implemented with an optically isolated monolithic semiconductor laser \cite{briles2020generating}, there are important constraints on the laser cavity to enable rapid frequency sweeping. Furthermore, optical isolators pose significant challenges for photonic integration, due to the compatibility of magneto-optic materials with silicon or III-V fabrication \cite{huang2018towards}.

Recently\cite{liang2015high, stern2018battery, pavlov2018narrow,raja2019electrically,voloshin2019dynamics,shen2020integrated}, a new approach to microresonator-soliton generation has emerged in which resonant Rayleigh back-scattering enables time-delayed coupling of a laser mode and microresonator mode. In this method, termed self-injection locking (SIL), a semiconductor laser is directly coupled to a microresonator chip, eliminating the need for an optical isolator and reducing the output power requirement of the laser, making it relatively compatible with hybrid integration.  Moreover, soliton generation with SIL exploits the dynamic properties of optical feedback to enable passive stabilization of the laser-frequency detuning with respect to the microresonator, simplifying the frequency sweep required for soliton generation. Specifically, the fast feedback outpaces microresonator thermal dynamics, therefore the laser frequency tracks the microresonator mode in order to maintain an appropriate pump field for the soliton microcomb.  

To date, this approach has been used for the generation of narrow-bandwidth soliton microcombs using multimode lasers, including reflective semiconductor optical amplifiers and SiN resonators \cite{stern2018battery}, as well as with Fabry-Perot (FP) semiconductor lasers with MgF$_2$ resonators \cite{pavlov2018narrow}, and with SiN resonators \cite{raja2019electrically}.  Soliton generation using SIL of single mode sources offer some advantages over multi-frequency sources because they are inherently less susceptible to complications resulting from mode-competition.  Recent SIL studies with single mode, distributed feedback (DFB) lasers have achieved narrow bandwidth solitons in MgF$_2$\cite{liang2015high} and SiN\cite{voloshin2019dynamics,shen2020integrated} resonators. 




In this work, we leverage recent experimental and theoretical work on SIL to create an octave-span, electrically-pumped soliton microcomb.  Our work utilizes hybrid integration of an InP DFB laser and a SiN resonator.  We compare the relative strengths of our method with alternative integration strategies based on direct current modulation of optically isolated DFB lasers \cite{briles2020generating} and spontaneous soliton formation in photonic crystal resonators (PhCR) \cite{yu2020spontaneous}.  In particular, analysis of the nonlinear SIL tuning curves \cite{voloshin2019dynamics} reveals a coupling between the accessible soliton detuning range and the strength of the linear scattering rate. Therefore, SIL microcombs exhibit a modified soliton stability diagram that significantly influences DW formation. Our work explores DW formation via simulations and measurements. Furthermore, in the course of our experiments we explored SIL with a FP laser instead of a DFB laser.  While the FP-laser SIL system would potentially mitigate some detuning limitations and simplify alignment of pump laser emission to SiN resonance compared to the DFB system, our experiments indicate that FP SIL microcombs are more challenging to reliably initiate, merely with electronic control of the FP laser injection current.  



\section{Comparison of hybrid integrated soliton pumping strategies
\label{sec:strategies}}

\begin{figure}[ht]
\centering
\includegraphics[width = 3.3in]{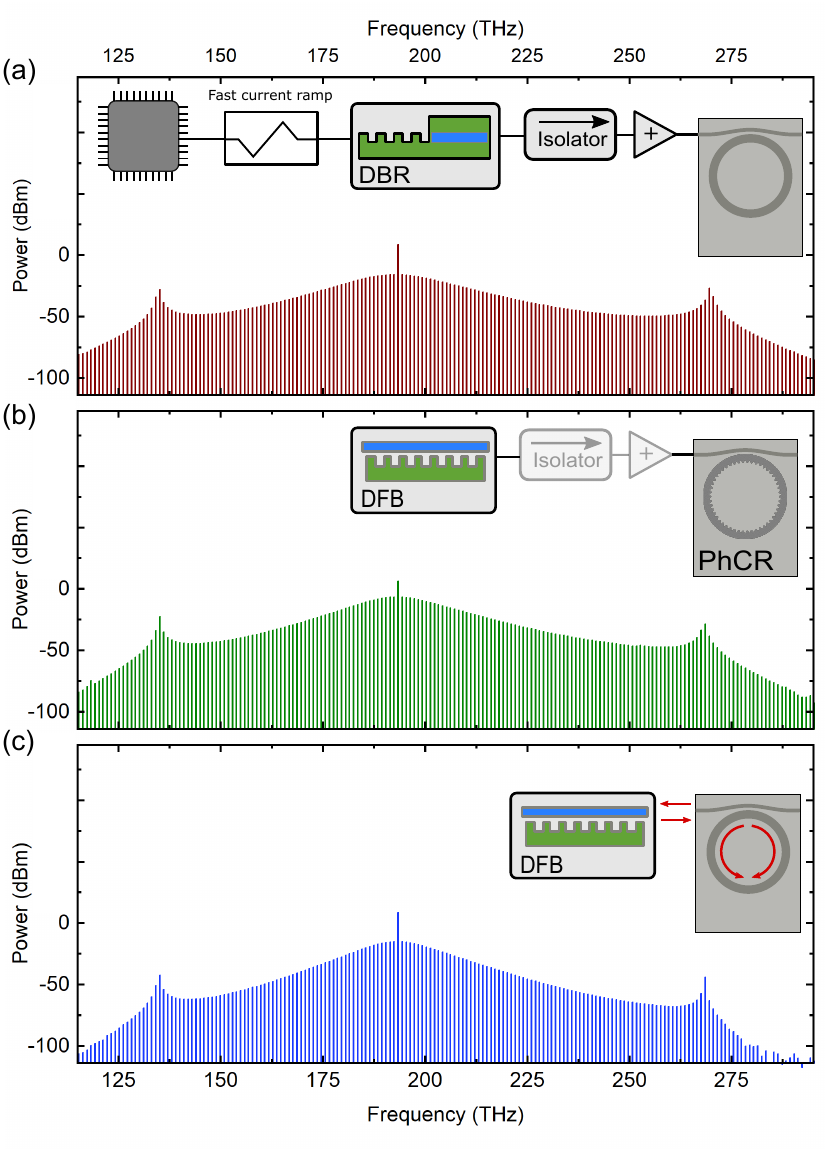}
\caption{\label{fig:Fig1} 
System configurations and LLE simulations to realize an octave-span soliton microcomb with a 1 THz line spacing. (a) Direct current modulation of an optically isolated, distributed-Bragg-reflector (DBR) with re-amplification to overcome losses.  The isolator enables arbitrary control of detuning to optimize soliton bandwidth and enter the intense DW regime. The simulated LLE spectrum corresponds to intrinsic quality factor, $Q_i = 2.5 \times 10^6$, coupling parameter, $K=1$, on-chip pump power $F^2 = 20$ (8 mW on-chip) and $\alpha = 24$. 
(b) Spontaneous soliton generation in a photonic crystal resonator (PhCR) pumped by a DFB laser.  The spectrum is simulated with $Q_i = 670\times 10^3$, $K=2$, $F^2 =1.44$ (13.7 mW on-chip), $\alpha = 1.56$ and pump-mode frequency shift of $\delta_{\text{PhCR}} = 1.54$. 
(c) Soliton generation via self-injection locking (SIL) of a DFB laser. The simulated spectrum has identical $Q_i$ and $K$ as (a) but introduces a normalized Rayleigh scattering rate $\Gamma = 0.95$ to provide resonant feedback to DFB.  Smaller values of $\alpha$ obtainable in the SIL regime limits the soliton bandwidth and DW power. $\alpha = 14.5$. 
} 
\end{figure}

Due to the progress and challenge of generating octave-span solitons suitable for $f$-$2f$ self-referencing, we analyze three plausible strategies that could realize an integrated soliton microcomb. The central consideration is how we affect control over the laser-microresonator frequency detuning, henceforth detuning. Figure \ref{fig:Fig1} presents an integration schematic and a simulation of the anticipated soliton-microcomb spectrum for each strategy.  The specific strategies are laser pumping with aid of an optical isolator to enable free manipulation of detuning (Fig. 1a), inclusion of a periodic nanostructure in the microresonator for enhanced control of nonlinear frequency shifts (Fig. 1b), and SIL of a DFB laser to the microresonator (Fig. 1c).  We envision that each microcomb chip is composed of a GVD-engineered resonator and its access waveguide, which utilizes an extended pulley-coupler segment to control external coupling across the 1--2 $\mu$m spectral range.

We base our analysis on numerical simulations with the Lugiato-Lefever equation (LLE) \cite{lugiato2018lugiato,coen2013modeling}, which offers a universal description of soliton behavior in terms of the resonator nonlinearity, dispersion, losses, pump laser power, and detuning. As a result, we identify a common resonator geometry, hence a common resonator GVD, which is compatible with generating an octave span comb.  Our model resonator has a 1 THz free-spectral range and supports what we refer to as a longwave dispersive wave (LDW) at frequency $f$ and a shortwave dispersive wave (SDW) at frequency $2f$.  We note that while harmonic (LDW$=2\times$ SDW) dispersive waves are advantageous for $f$--$2f$ detection of $f_{\text{ceo}}$, such enhancements are not strictly required for self-referencing.  The specific dispersion profile is derived from finite element simulations (\emph{COMSOL Multiphysics}\footnote{Mention of specific companies or trade names is for scientific communication only, and does not constitute an endorsement by NIST.}) of the resonator mode frequencies for an oxide-clad geometry with resonator waveguide width (RW) and silicon nitride film thickness of 1742 nm and 780 nm, respectively. We anticipate that this configuration will result in an LDW at 135 THz (2220 nm) and an SDW at 270 THz (1110 nm) with a pump-laser frequency of $\omega/2\pi$ =193 THz (1550 nm). 

Aside from a common resonator GVD to control the LDW and SDW, we use a common framework to describe the laser power and detuning for accurate comparisons. We normalize the pump power in the access waveguide, $P_{\text{in}}$, to the threshold power for parametric oscillation ($P_{\text{th}}$) according to $F^2 = P_{\text{in}}/P_{\text{th}}$. In particular, we constrain our analysis to $F^2$ values consistent with chip lasers.  The output coupling rate of the microresonator to the access waveguide is also mixed into the required $F^2$ setting.  The coupling parameter is $K =\kappa_c/\kappa_i = Q_i/Q_c$, where $\kappa_c$ is the coupling loss rate, $\kappa_i$ is the intrinsic loss rate, and $Q_i = \omega_0/\kappa_i$ and $Q_c = \omega_0/\kappa_c$ are the associated quality factors at the angular resonant frequency $\omega_0$. We define the detuning parameter in the LLE as $\alpha= 2(\omega_0-\omega)/\kappa$ where $\kappa=\kappa_i+\kappa_c$ is the total loss rate such that $\alpha>0$ corresponds the pump laser red detuned from the resonator mode.  Solitons only form at certain combinations of $F^2$ and $\alpha$ \cite{godey2014stability}.  The power-dependent, minimum and maximum detuning of soliton stability is estimated by $\alpha_\text{min} = (F/2)^{2/3}+\sqrt{4 \left(F/2 \right)^{4/3}-1}$ and $\alpha_{\text{max}} = \pi^2 F^2/8$, respectively \cite{godey2014stability,herr2014temporal,voloshin2019dynamics}. A critical aspect of experimental work with soliton microcombs is that power-dependent thermal shifts of $\omega_0$ and $\alpha$ cannot generally be inferred from straightforwardly measurable system parameters, e.g. $\omega_0$ at low laser power and the laser frequency.

We first consider soliton generation by rapid injection current sweeps of an optically isolated, single-frequency semiconductor laser; see Fig. \ref{fig:Fig1}(a). An appropriate current sweep can achieve a laser frequency sweep that suppresses thermal shifts in $\omega_0$, leading to dynamic stabilization of $\alpha$ in the range $[\alpha_{\text{min}},\alpha_{\text{max}}]$. Experience with distributed-Bragg-reflector (DBR) lasers in the 1064 nm band \cite{briles2020generating} and DFB lasers in the 1550 nm band \cite{nishimoto2020investigation} confirm this approach. This method's principal strength is unrestricted detuning control between $\alpha_{\text{min}}$ and $\alpha_{\text{max}}$, which enables us to optimize detuning for maximum optical power in the SDW and LDW.  The strongest DWs are typically observed for $\alpha \approx \alpha_{\text{max}}$.  In particular, using the LLE to calculate the soliton spectrum of our model resonator, we obtain the spectrum in Fig. 1a with $\alpha = 24 \approx 0.97\times \alpha_{\text{max}}$.  Here we utilize critical coupling ($K=1$) and an internal quality factor, $Q_\text{i}= 2.5\times 10^6$, which results in $P_{\text{th}}= 0.4$ mW; $F^2=20$ corresponds to an on-chip power of only 8 mW. The LDW and SDW powers that we obtain are -27.5 dBm and -26.5 dBm, respectively. These conditions reflect a compromise between maximizing DW power, which requires a setting of $\alpha$ near the $\alpha_{\text{max}}$ boundary, \cite{briles2020generating} and suppressing complex, fascinating, and not yet fully understood soliton breathing oscillations that have been experimentally \cite{briles2017kerr,cole2019subharmonic} and theoretically \cite{skryabin2017self} described.

In Fig. \ref{fig:Fig1}(b), we explore the recent innovation of photonic-crystal resonator (PhCR) soliton microcombs \cite{yu2020spontaneous}, which are generated without complicated detuning control and our findings here indicate that they can enable intense soliton DWs. The PhCR is composed of a lithographically defined periodic modulation of a conventional ring resonator. The PhCR bandgap introduces a controllable resonant frequency shift of one (or potentially more) azimuthal modes. By programming this frequency shift to balance the Kerr shift of the pump-laser mode, we enable spontaneous soliton generation with $\alpha$ much closer to zero than with a conventional microresonator. We define the normalized frequency shift as $\delta_{\text{PhCR}}=\left(\omega_0-\omega_0'\right)/\kappa$ where $\omega_0'$ is the angular frequency of the pump mode in the absence of the PhCR. Operationally, the magnitude of the periodic modulation controls $\delta_{\text{PhCR}}$, and we assume negligible effect on the model GVD design. We assume overcoupling ($K=2$) and $Q_i = 675 \times 10^3$, which results in  $P_{\text{th}} = 9.3$ mW.  These conditions reflect a balance between the optimum conditions for spontaneous solitons and high output coupling of comb power.  The pump power is $F^2 = 1.44$ (13.4 mW on-chip), and we choose $\delta_{\text{PhCR}}=1.54$ and $\alpha = 1.56$ to ensure preferential generation of a single soliton with LDW and SDW powers of -22 dBm and -27.5 dBm, respectively. This PhCR design provides noticeably higher DW power than a conventional microcomb would provide at a comparable $F^2$ pump power. Existing experiments have not explored hybrid integration of a laser and PhCR.

\begin{figure*}[t] \centering
\includegraphics[width = 0.90\textwidth]{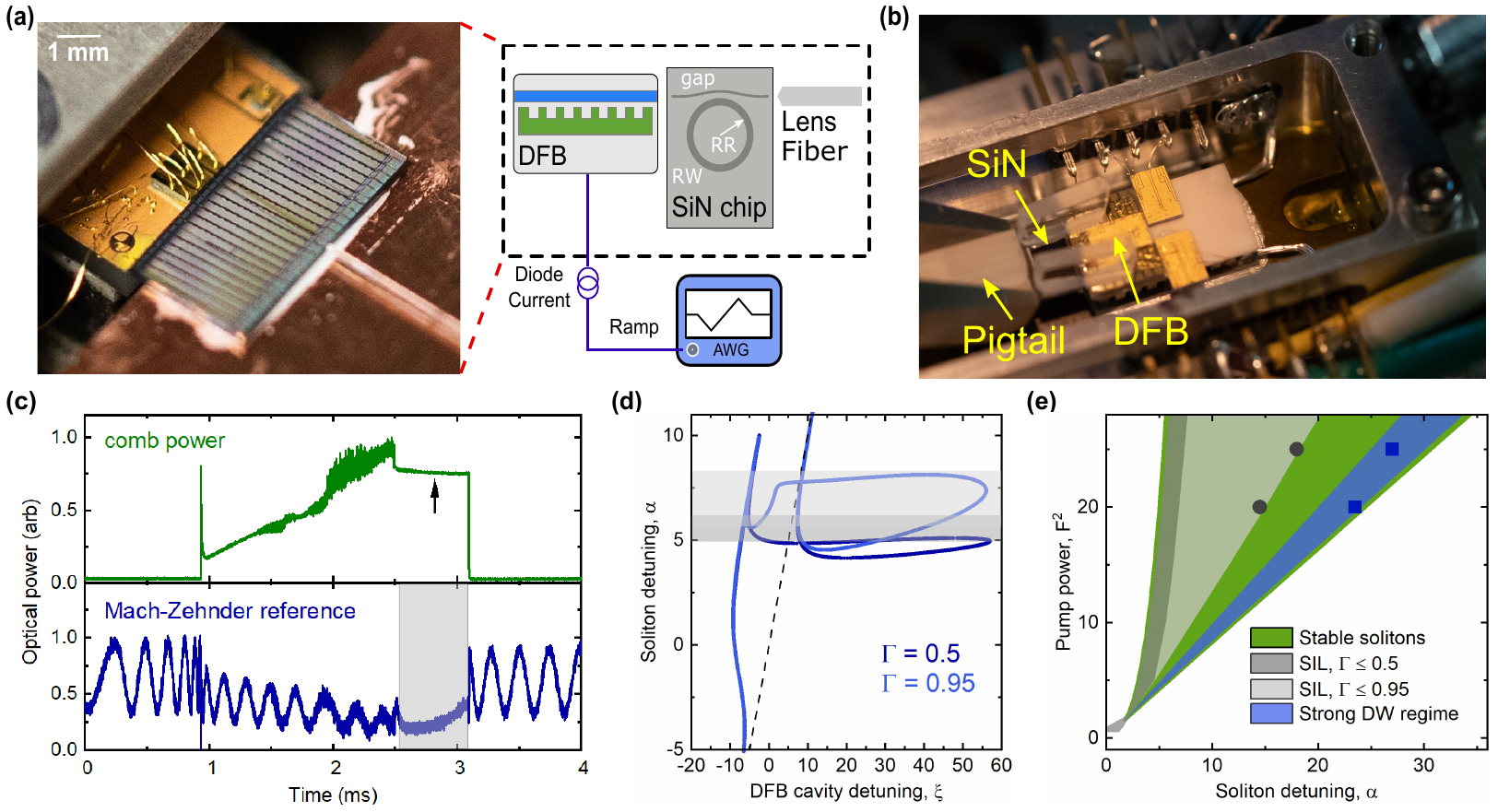}
\caption{\label{fig:Fig2} 
(a) Photo and diagram highlighting the  key components for benchtop SIL experiments.  A chip-scale DFB laser is butt-coupled to a SiN chip containing a microresonator where Kerr solitons are generated and coupled off-chip with a lensed fiber.  AWG, arbitrary waveform generator. (b) Hybrid-integrated soliton microcomb where components in (a) are mounted into a single butterfly package.  Note that in (b), the order of the components from left to right is opposite to that in (a).  (c) Oscilloscope trace of the photodetected comb power (top panel) and DFB laser frequency (bottom panel), monitored with a Mach-Zehnder interferometer, during a laser-current sweep.  The SIL region (gray rectangle) overlaps with a soliton step marked by a black arrow.  (d) Nonlinear tuning curves for SIL for normalized scattering rates, $\Gamma=0.5$ (dark blue) and $\Gamma=0.95$ (light blue), for fixed $F^2 = 10$, locking coefficient, $K_{\text{SIL}} = 250$ and feedback phase $\psi_0=0$.  The horizontal axis is the detuning between the DFB laser and the microresonator mode.  The horizontal axis is the normalized detuning between the lasing DFB cavity mode frequency and the pumped microresonator mode frequency, $\xi$.  
The vertical axis is the soliton detuning, $\alpha$.  The range of accessible $\alpha$-values in the SIL regime for $\Gamma = 0.5$ ($\Gamma=0.95$) are shown as gray (light gray) horizontal bars; note the increased range for $\Gamma=0.95$. (e) An ($\alpha,F^2$) soliton diagram, indicating the soliton stability regime (green shaded region), the strong DW regime (blue region), and the SIL regime for $\Gamma \leq 0.5$ (dark gray region) and $\Gamma \leq 0.95$ (light grey region). The green soliton region contains both SIL regimes and the strong DW regime.  Simulation parameters used in Figs. \ref{fig:Fig1}(a) and \ref{fig:Fig1}(c) ($F^2=20$) and the middle and bottom panel of Fig. \ref{fig:Fig3}(a) ($F^2=25$) are indicated with blue squares and black circles for the for the strong DW and SIL regimes respectively.
}\end{figure*} 

The third case we analyze is a SIL soliton microcomb with a DFB laser.  Similar to the PhCR strategy (Fig. \ref{fig:Fig1}(b)), this approach doesn't require complicated laser frequency sweeping, and it has the advantage of being compatible with a conventional microresonator.  This system relies on Rayleigh backscattering, which breaks the degeneracy of forward and backward propagation in the microresonator, leading to distinct resonance peaks.  The visibility of this splitting is described by the normalized scattering rate $\Gamma = \beta/\kappa$, where $\beta$ is the linear scattering rate.  For $\Gamma\ll1$, no observable scattering doublet is observed but as $\Gamma$ increases, two distinct resonance peaks can be resolved with separation $\beta$.  Figure \ref{fig:Fig1}(c) presents the expected SIL microcomb spectrum, highlighting the spectral bandwidth and DW power.  The modeled spectrum has identical resonator loss rates and on-chip pump powers as the conventional microresonator case (Fig. \ref{fig:Fig1}(a)), namely $Q_i = 2.5 \times 10^6$, $K=1$, and $F^2=20$.   We additionally include a normalized scattering rate of $\Gamma \approx 1$ in correspondence with typical experimental data; see Fig. \ref{fig:Fig2} and \ref{fig:Fig3}. 
As described in Sec \ref{sec:SecIII} and Fig. \ref{fig:Fig2}, SIL microcombs operate in a restricted range of $\alpha$.  However, within this restricted range, there is an $F^2$-dependent coupling between $\Gamma$ and $\alpha$ such that the $\alpha$ range modestly increases with $\Gamma$.  The spectrum in Fig. \ref{fig:Fig1}(c) corresponds to $\alpha = 14.5$, the maximum soliton detuning that can be accessed at $F^2=20$ and $\Gamma=0.95$. This reduced value of $\alpha$ compared to Fig. \ref{fig:Fig1}(a) leads to a more than 15 dB reduction in the power of both DWs (-42 dBm and -44 dBm for the the LDW and SDW respectively).  Such a significant reduction in DW powers inevitably would lead to a reduction in the signal-to-noise ratio of $f_\text{ceo}$ detection with an $f$-$2f$ interferometer.

\section{Hybrid Integrated Combs and Accessible Soliton Detunings
\label{sec:SecIII}}

In light of the above analysis, we turn to constructing a hybrid-integrated, octave-span soliton microcomb enabled by SIL. We present our experimental setups and characterization measurements, which we analyze by applying previously derived SIL dynamics \cite{voloshin2019dynamics} to the case of an octave-span soliton microcomb. These data allow us to evaluate our observations, particularly with respect to thermal stabilization of the microresonator during soliton formation and the range of accessible detuning that control DW intensity.

Figures \ref{fig:Fig2}(a) and \ref{fig:Fig2}(b) show our hybrid integrated microcomb systems, which contain a DFB laser, a SiN microresonator chip, and a lensed fiber to collect the soliton microcomb output.  Our experiments utilize SiN chips that contain approximately 400 resonators with independent access waveguides. The SiN devices were fabricated by Ligentec\cite{NISTcommercialSTATEMENT}, yielding approximately 100 chips. This high density of resonators on a single $3 \times 5$ mm chip allows for wide variation in resonator and access waveguide coupling geometries to support ultralow-power, octave-span solitons \cite{briles2020generating}.  We use pulley couplers \cite{moille2019broadband}, and we control $K$ by varying the length of the pulley coupler $L_c$ and the waveguide-resonator gap, $G$.  Our SiN devices have an average $Q_i = (2.70 \pm 0.54)\times 10^6$, $L_c = 17\ \mu$m and $G = 700$ nm or $G = 800$ nm.  The typical threshold power for parametric oscillation in these devices is 1.5 mW. 

We use a InP laser chip with four DFB laser that is mounted to a butterfly package and positioned relative to the SiN with a multi-axis stage. We butt couple \cite{boust2020microcomb} the InP laser chip and the SiN chip with a gap of approximately 3 $\mu$m. Inverse taper couplers at the SiN chip edge enable 5.5 dB and 3 dB insertion loss to the DFB laser and lensed fiber, respectively. We use the setup in Fig. \ref{fig:Fig2}(a) for screening and modular testing. Fig. \ref{fig:Fig2}(b) shows a photo of our fully integrated, electrically-driven microcomb source, consisting of a DFB laser, a SiN resonator chip and bonded output fiber pigtail, permanently mounted in a single butterfly package. We mount the laser and SiN chips on independent copper tungsten heat sinks and align them with a precision positioning system. Glass trusses bonded across the top of the assembly with UV epoxy fix the gap between the laser chip and the SiN chip. We monitor soliton formation continuously (as in Fig. \ref{fig:Fig2}(c)) until the epoxy has completely cured.  To mitigate misalignment from temperature gradients, we maintain a common substrate temperature of the laser and SiN with a TEC underneath the butterfly package.  This architecture choice eliminates the possibility of independent temperature control except for microheaters.  




To operate the SIL microcomb system, we scan the injection current of the DFB laser with a signal generator, causing the laser frequency to scan across a ring resonator mode. An essential requirement of packaging experiments with the DFB laser SIL microcomb is frequency alignment of the laser and ring resonator. Since the DFB laser tuning range is a relatively small $1556 \pm 1$ nm, we rely on discrete adjustments of the SiN ring radius ($RR$) in steps of 100 nm, obtained by changing to different devices on the same chip, to adjust the resonator frequency into the DFB laser's range. These $RR$ variations also cause an undesirable variation of $f_\textrm{ceo}$ that would need to be mitigated by some other element of a phase-stabilized SIL microcomb system. Conversely, the DW wavelengths are relatively insensitive to RR, hence we are able to generate a consistent output spectrum.

We characterize SIL of the DFB laser to the microresonator and soliton formation by monitoring the output from the SiN chip that couples to the lensed fiber. We monitor the microcomb's optical power, which we derive by photodetection after wavelength-selective filtering to attenuate the pump power, and the pump-laser frequency, which we deduce from photodetection of fringes after transmission of a Mach-Zehnder (MZ) interferometer with a free-spectral range of 570 MHz. These quantities are proportional to the photocurrent and the period of MZ fringes, respectively. Figure \ref{fig:Fig2}(c) indicates the evolution of comb power (top panel) and pump frequency (bottom panel) as we scan the DFB-laser injection current to vary the frequency from high to low with respect to the microresonator mode; the frequency scan range is sufficiently wide to access all observable nonlinear behaviors. Specific features of these data are directly diagnostic of our system. The comb power increases in a manner\cite{cole2019subharmonic} consistent with spatiotemporal instabilities of the Turing pattern that arises at the threshold for parametric oscillation \cite{coillet2019transition}. Specifically, comb-power noise indicates a chaotic intraresonator waveform. We observe stable soliton formation consistent with a sufficiently negative detuning, signified by the abrupt change in comb power and reduction in comb-power noise; see the arrow in Fig. \ref{fig:Fig2}(c). Importantly, in this experiment we also observe SIL that coincides with soliton formation, signified by a complete reduction in the rate of MZ fringes.  To optimize for this condition, we vary the feedback phase by altering the gap between the DFB and SiN resonator chip.  Thermal bistability is apparently also more manageable in SIL soliton formation. We observe the soliton for a duration 0.5 ms, which is substantially longer than the $\sim100$ ns thermal conduction timescale of the resonator, even during the DFB-laser frequency scan. Hence, SIL plays a critical role in dynamically maintaining the detuning required to energize the soliton, making this approach important for some applications of microcombs.

\begin{figure*}[t!]
  \centering
\includegraphics[width = 6.8in]{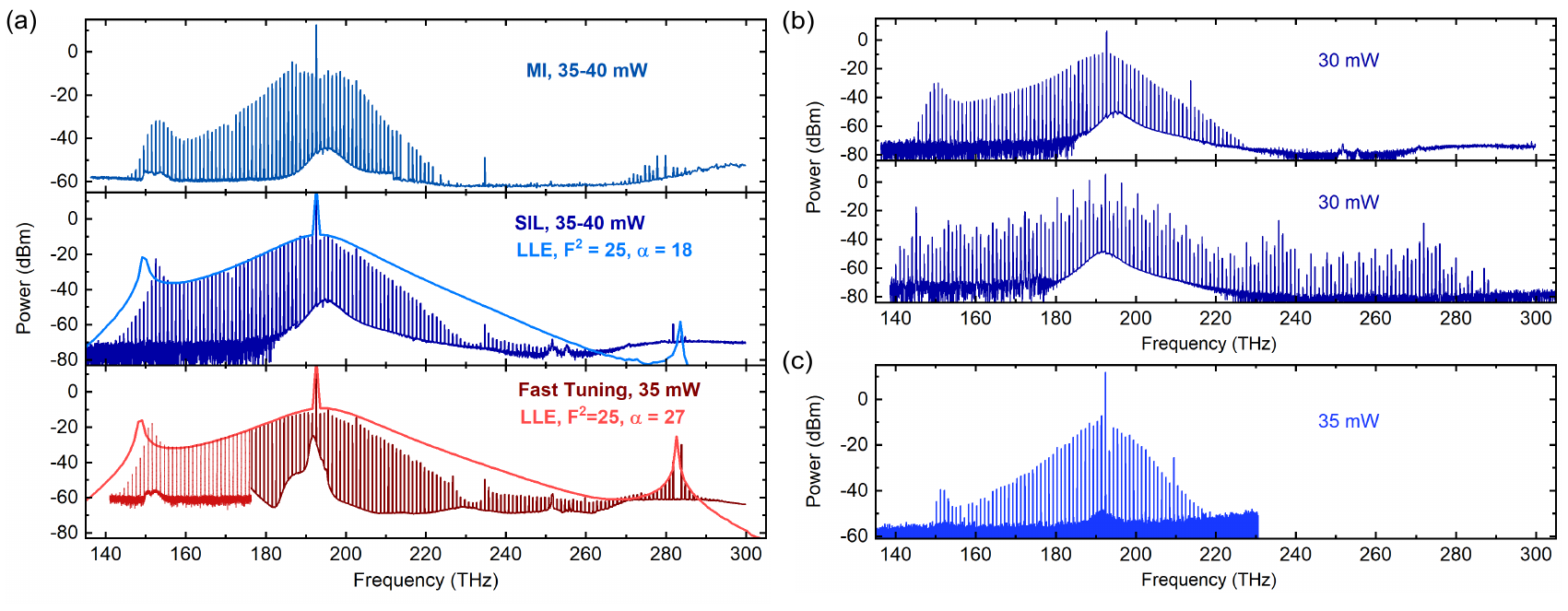}
\caption{\label{fig:Fig3}  
Exploring self-injection-locked soliton microcomb spectra. (a) Comparison of experimental spectra obtained under different conditions, but using the same resonator with a RW of 1653 nm and similar on-chip power of 35--40 mW. Using the butt-coupled DFB laser and SiN chip of Fig. 2(a), we obtain the top spectrum (lighter blue) of an MI microcomb pattern and the middle spectrum (darker blue) of a SIL soliton microcomb. For comparison, the bottom spectrum (red) shows a soliton microcomb obtained with an optically isolated laser as in Fig. 1(a) that enables a 30 dB more intense SDW.  The overlaid curves of the middle and bottom panel are soliton spectral envelopes from LLE simulations under the conditions of maximum accessible detuning at $F^2 \approx 25$ in the SIL regime ($\alpha = 18$, light blue) and a larger detuning in the intense DW regime ($\alpha = 27$, light red).  These results highlight the accuracy of LLE modeling. 
(b) SIL soliton microcombs adjusted to obtain different soliton number. Top, single soliton spectrum utilizes a RW of 1620 nm and an on-chip power of 30 mW. Bottom, multiple soliton spectrum utilizes a RW of 1642 nm and an on-chip power of 30 mW.
(c) Optical spectrum of the soliton generated with the fully packaged hybrid-integrated device of Fig. 2(b) with 35 mW of on-chip power.
} 
\end{figure*}

We perform numerical modeling, according to Ref. \cite{voloshin2019dynamics} to understand dynamic thermal stability and the range of accessible detuning in SIL soliton microcombs. The model describes the DFB laser frequency $\omega$ under the influence of Rayleigh backscattering from the microresonator. Without nonlinearity, a SIL laser-microresonator system should operate at a detuning outside the range needed for soliton formation \cite{kondratiev2017self}. In Kerr-nonlinear microresonators, self- and cross-phase modulation between the clockwise and counter-clockwise pump fields leads to an additional nonlinear detuning shift, $\alpha_{\text{NL}}$ and coupling shift, $\Gamma_{\text{NL}}$. Therefore, the SIL laser-microresonator operates at red detuning, favoring soliton formation \cite{shen2020integrated, voloshin2019dynamics}. This feature combined with the enhanced, passive thermal stability described above, have been exploited to create turnkey Kerr comb systems \cite{shen2020integrated}.  However, these previous studies did not consider what value of $\alpha$ could be reached in the SIL regime and what effect this has on the soliton spectrum.

Here, we describe a coupling between $\Gamma$ and $\alpha$, and we discuss the implications for our octave-span SIL microcombs. The parameters $\alpha_{\text{NL}}$ and $\Gamma_{\text{NL}}$ are strongly dependent on $F^2$ and coupled together through the following system of equations \cite{voloshin2019dynamics},

\begin{equation}
\alpha_{\text{NL}} = \frac{2\theta+1}{2} F^2 \frac{1 + \left(\bar{\alpha}-\Gamma_{\text{NL}}\right)^2 + \Gamma^2}{\left(1 + \bar{\Gamma}^2 - \bar{\alpha}^2 \right)^2 + 4 \bar{\alpha}^2}
\label{Eq:alphaNL}
\end{equation}
and
\begin{equation}
\Gamma_{\text{NL}} = \frac{2\theta-1}{2} F^2 \frac{1 + \left(\bar{\alpha}-\Gamma_{\text{NL}}\right)^2 - \Gamma^2}{\left(1 + \bar{\Gamma}^2 - \bar{\alpha}^2 \right)^2 + 4 \bar{\alpha}^2}
\label{Eq:gammaNL}
\end{equation}
\noindent where $\bar{\alpha} = \alpha - \alpha_{\text{NL}}$, $\bar{\Gamma}^2 = \Gamma^2 + \Gamma_{\text{NL}}^2$ and $\theta$ is a mode overlap factor assumed to be unity.  We numerically solve Eqns. \ref{Eq:alphaNL} and \ref{Eq:gammaNL} for chosen values of $F^2$, $\Gamma$ and $\alpha$.  This is repeated for different values of $\alpha$ and collected into arrays to generate the detuning dependence of $\alpha_{\text{NL}}$ and $\Gamma_{\text{NL}}$ from which we calculate the normalized intracavity power for the forward mode ($|a|^2$) and backward mode ($|b|^2$), using
\begin{equation}
|a|^2 = \frac{\alpha_{\text{NL}}}{3} + \Gamma_{\text{NL}}
\label{eq:forwardPOWER}
\end{equation}
and
\begin{equation}
|b|^2 = \frac{\alpha_{\text{NL}}}{3} - \Gamma_{\text{NL}}.
\label{eq:backwardPOWER}
\end{equation}


\noindent In Eq. \ref{eq:forwardPOWER} and Eq. \ref{eq:backwardPOWER}, we have assumed $\theta = 1$.  The DFB cavity mode frequency, $\omega_d$, is described in terms of the normalized detuning ($\xi$) between the pumped microresonator mode, $\omega_0$ which we define as $\xi = 2\left(\omega_0-\omega_\text{d}\right)/\kappa$.  Finally, the full nonlinear tuning response of $\xi$ as a function of $\alpha$ is given by

\begin{equation}
    \xi =   \alpha + \frac{K_{\text{SIL}}}{2} \frac{2 \bar{\alpha} \cos \bar{\psi}+ \left(1 + \bar{\Gamma}^2 - \bar{\alpha}^2\right) \sin \bar{\psi} }{ \left(1 + \bar{\Gamma}^2 - \bar{\alpha}^2 \right)^2 + 4 \bar{\alpha}^2}
    \label{eq:nonlinearTUNINGcurve}
\end{equation}

\noindent In Eq. \ref{eq:nonlinearTUNINGcurve}, $K_{\text{SIL}} = 8 \eta \kappa_{\text{do}} \sqrt{1 + \alpha_g^2}/(\kappa R_0)$ is the locking strength of the coupled microresonator-DFB system, where $\alpha_g$ is the Henry factor, $R_0$ is the DFB facet reflectivity and $\kappa_{\text{do}}$ is the output mirror loss rate.   $\bar{\psi} = \psi_0 - \kappa \tau_s \alpha/2 $ is the detuning dependent feedback phase where $\tau_s$ is the round trip-time between the DFB facet and microresonator coupler.  

Figure \ref{fig:Fig2}(d) shows two tuning curves of $\alpha$ as a function of $\xi$ with $\Gamma = 0.5$ (dark blue) and $\Gamma = 0.95$ (light blue) for $F^2 = 10$ and $K_{\text{SIL}}=250$.  Experimentally, $\xi$ is the parameter that we can control with the DFB injection current, $I_\text{inj}$, which directly controls $\omega_{\text{d}}$.  In the absence of optical feedback, $\omega$ follows the change in $\omega_{\text{d}}$; see dashed line corresponding to $\alpha=\xi$ in Fig. \ref{fig:Fig2}(d).  Under these conditions the emission frequency of our DFB laser tunes linearly with laser current, with a tuning coefficient of $\gamma = \frac{1}{2\pi}\frac{d\omega}{dI_{\text{inj}}} \approx 0.7$ GHz/mA.  In the presence of resonant Rayleigh-scattering, the tuning curve becomes distorted due to frequency pulling. As in the case of SIL in linear resonators \cite{kondratiev2017self}, this optical feedback leads to a reduction in $\gamma$ (gray rectangle in Fig. \ref{fig:Fig2}(c)).  This hallmark of SIL is manifested in the tuning curves as a locking plateau where $\alpha$ is insensitive to changes in $\xi$.

We are principally interested in the range of $\alpha$ that are accessible in the SIL regime, $\alpha_{\text{SIL}}$, instead of the associated $\xi$ capture range.  We determine this range by analyzing stable branches of the tuning curve and the associated forward intracavity power curve (Eq. \ref{eq:forwardPOWER}).  The locking plateau becomes accessible for $\psi_0 \approx 0$, which is controlled experimentally by varying the physical gap between the DFB and the SiN resonator. Altering the phase about this favorable locking condition can also be used to alter the detuning, and we used this fact to maximize the soliton bandwidth \cite{shen2020integrated}.  As $K_{\text{SIL}}$ governs the overall distortion of the tuning curve and slope of the locking plateau, larger values of $K_{\text{SIL}}$ may be required in order for access to the locking plateau.  The soliton detuning range increases considerably for $\Gamma =0.95$ (light gray band) compared to $\Gamma=0.5$ (dark gray). 

Figure \ref{fig:Fig2} (e) summarizes the available pairings of $F^2$ and $\alpha$ for a few settings of $\Gamma$ with a backdrop of the soliton existence range [$\alpha_\textrm{min}$, $\alpha_\textrm{max}$].  This information encapsulates many properties of soliton microcombs, especially their optical spectra. For small values of $\Gamma<0.5$, $\alpha_{\text{SIL}}$ follows the shape of the minimum detuning boundary $\alpha_{min}$ with a roughly uniform width.  For $\Gamma >0.5$, the maximum value of $\alpha_{\text{SIL}}$ begins to increase significantly with both $\Gamma$ and $F^2$.  Our analysis is limited to $\Gamma < 0.95$, as instabilities can arise for larger $\Gamma$-values \cite{kondratiev2020modulational} values in some cases.  To place our earlier simulations of octave-span microcombs in perspective, in Fig. \ref{fig:Fig2}(e) we have marked the LLE parameters used in our SIL simulations, shown in Fig. \ref{fig:Fig1}(c) and in the middle panel of Fig. \ref{fig:Fig3}(a), with black circles.   $\alpha_{\text{SIL}}$ does not quite reach the detunings typically required to trigger intense dispersive waves, which we have indicated by the blue region in Fig. \ref{fig:Fig2}(e).  Simulation parameters in the strong DW regime from Fig. \ref{fig:Fig1}(a) and the bottom panel Fig. \ref{fig:Fig3}(a) are marked with blue squares.  


\section{Octave-span SIL microcombs}

Figure \ref{fig:Fig3} presents experimental SIL stable soliton spectra generated with our benchtop and hybrid-integrated systems.  In Fig. \ref{fig:Fig3}(a), we compare SIL using our benchtop system to an optically isolated pumping strategy, and we analyze the consequences of the modified stability diagram presented in Fig. \ref{fig:Fig2}(e), using LLE simulations.  All spectra in Fig. \ref{fig:Fig3}(a) were generated with 35-40 mW of on-chip power in the same resonator with nominal RW = 1653 nm.  We achieve conventional pumping of a modulation instability (MI) comb by tilting the DFB relative to the SiN chip to prevent optical feedback from reaching the DFB.  Reduction of this angular misalignment allows the SIL regime to be accessed, generating a near-octave span single soliton with LDW and SDW powers of -21 dBm and -60 dBm respectively (dark blue, middle panel).  This SIL spectrum is markedly different than that obtained in the same resonator using fast frequency ramps from an optically isolated laser system (bottom panel, dark red). Due to availability of measurements, we have supplemented the shortwave spectrum from 175-300 THz (dark red) with the long-wave spectrum from a different resonator with identical $RW$, gap and thickness, but a ring radius that differed by 0.8 $\%$ (light red).  Such a small $RR$ variation has a negligible effect on the resonator GVD.  In contrast to the SIL case, the fast tuning method offers complete control of the laser pump frequency, ensuring that detunings within the intense DW regime can be reached.  Indeed, adjustment of $\alpha$ after generating the soliton results in a nearly 30 dB increase in SDW power near 283 THz (1059 nm).  Our experimental findings are supported by LLE simulations of the soliton spectral envelope in the SIL regime and strong DW regime which we have overlaid in the middle and bottom panel of Fig. \ref{fig:Fig3}(a) in light blue and light red, respectively.  The LLE simulations have the same normalized pump power as the experimental spectrum ($F^2=25$) and the maximum accessible detuning for $\Gamma = 0.95$ determined from Fig. \ref{fig:Fig2}(e).  For the SIL and strong DW regimes we use $\alpha =18$ and $\alpha=27$ respectively.  While there are discrepancies between the experimental and simulated spectrum between 200-250 THz due to wavelength dependent variation in $\kappa_\text{c}$ of our pulley coupler \cite{moille2019broadband}, there is reasonable agreement in the SDW powers.

\begin{figure}[htbp]
\centering
\includegraphics[width = 3.25in]{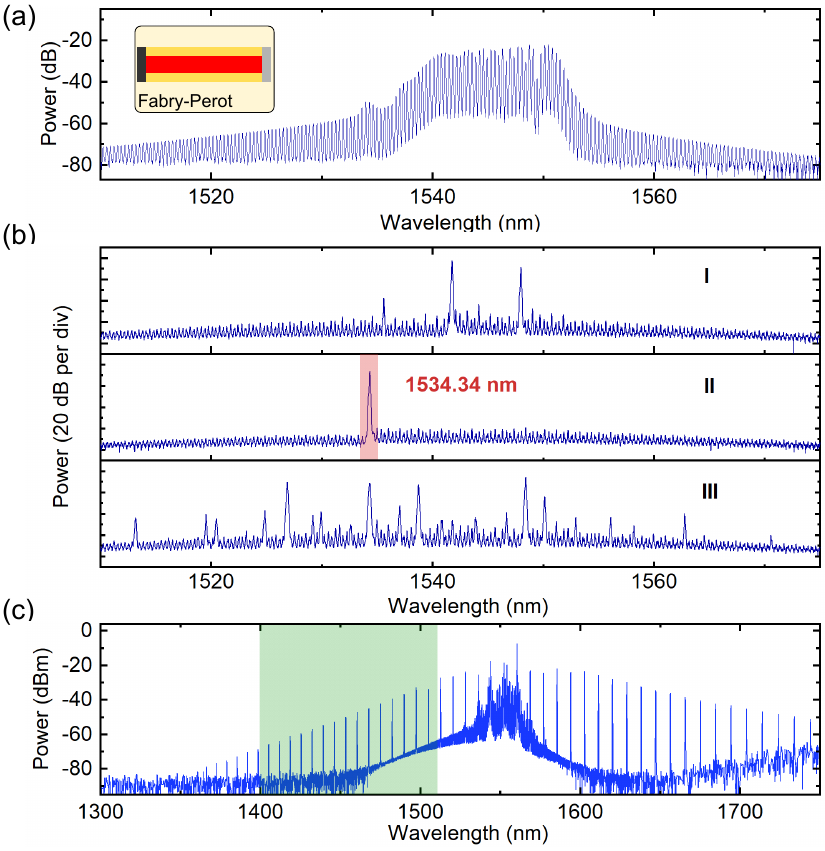}
\caption{\label{fig:Fig4} 
SIL with a FP semiconductor laser. (a) Diagram of a FP laser and the emission spectrum in absence of optical feedback.   (b) Emission spectrum with optical feedback.  Laser current is increased from top to bottom (state I to III).  State II shows SIL to one of the SiN resonator modes at 1534.34 nm while states I and III are influenced by reflections from SiN chip facet. The red shaded region indicates the spectral range focus of our diagnostic measurement system.  (c) MI microcomb generation with the FP laser. The green shaded region indicates the spectral range focus of our diagnostic measurement system.  Due to the small range of suitable currents, we did not observe reliable soliton formation with the full suite of diagnostic tools described in this paper.} 
\end{figure}

Figure \ref{fig:Fig3}(b) presents SIL soliton spectra obtained with our benchtop SIL system, using different resonator geometries with similar $Q_i$ and $\Gamma$.  The top panel shows a single soliton from the same SiN chip as Fig. \ref{fig:Fig3}(a) with nominal $RW$ = 1620 nm.  Similar to the spectra shown in the middle panel of Fig. \ref{fig:Fig3} (a), the SDW power suffers from the restricted values of $\alpha$ in the SIL regime.  The bottom panel shows a multi-soliton generated with a different resonator ($RW = 1642$ nm) with a smaller gap ($G= 700$ nm).  Here, more favorable broadband performance of the pulley coupler, combined with constructive spectral interference of the circulating solitons, yields higher DW comb power.   Figure \ref{fig:Fig3}(c) presents a  single-soliton spectra generated with our fully integrated system using 35-40 mW of power (Fig. \ref{fig:Fig2}(b)).  The resonator is from a different SiN chip as Fig. \ref{fig:Fig3}(a) but has an identical geometry.  
However, as in Fig. \ref{fig:Fig3}(a), the SDW power in the SIL suffers considerably.  We attribute this to two factors that restricted the soliton detuning, namely the fixed gap and inability to control soliton detuning with optical phase, and the overall restriction of obtainable detunings in the SIL state.

\section{Self Injection Locking with Fabry-Perot Lasers}

In the course of our experiments with a SIL soliton microcombs, we also investigated soliton generation with an injection locked, multi-frequency Fabry-Perot (FP) laser \cite{pavlov2018narrow,raja2019electrically}.  We found this approach to have significant drawbacks compared to SIL with single frequency DFB lasers.  For the aid of future researchers, we offer a description of some of these challenges; see Fig. \ref{fig:Fig4}.  While SIL strategies with multi-frequency sources such as FP lasers benefit from simple, low-cost components, this comes at the cost of higher operational complexity because the pump source must be converted to single frequency operation in order to pump the soliton.  Single frequency sources such as distributed feedback (DFB) lasers on the other hand, avoid mode competition effects and can be injection-locked over a wider range of injection currents \cite{savchenkov2019self}.  

In our experiments, we mounted a commercial FP laser chip into a butterfly package and butt-coupled it to a SiN chip, using an identical setup to that used for DFB injection locking experiments; see Fig. \ref{fig:Fig2}(a).  The free-running emission spectra of our FP laser is composed of discrete spectral lines spaced by FSR = 43.35 GHz that are centered around a 10 nm wide pedestal; see Fig. \ref{fig:Fig4}(a).  The pedestal bandwidth exceeds the SiN resonator 1 THz FSR (approximately 8 nm at 1550 nm) and we expected that this high density of modes would simplify frequency alignment between the pump laser and the resonator mode.  The presence of broadband emission from the FP laser that overlaps with expected soliton comb lines necessitates modified SIL diagnostics compared to DFB injection.  We diagnose the transition from multimode to single-mode operation by monitoring the laser power at the wavelength of the SiN resonator SIL target mode, using a narrow tunable bandpass; see red rectangle in middle panel of Fig. \ref{fig:Fig4}(b)).  Soliton formation is diagnosed by detecting laser power outside of the emission band of the FP laser; green rectangle of bottom panel of Fig. \ref{fig:Fig4}(c)).



Figure \ref{fig:Fig4}(b) analyzes the effect of optical feedback on the FP laser's output spectrum.  While both DFB and FP systems are susceptible to unwanted feedback from the SiN chip facets, it is especially problematic for FP lasers.  The presence of a small reflection from the facet results in an additional resonance condition in the coupled cavity dynamics for the laser-microresonator system.  For an FP laser, this effect is magnified by the presence of multiple lasing modes, each of which has a different phase condition. When the FP laser is butt coupled to the SiN chip, optical feedback from Rayleigh scattering and the SiN chip facets increases the gain of select modes still in a multi-mode regime (state I, top panel).  An increase in the diode current induces injection-locking to the SiN resonator mode (state II, middle panel).  Surprisingly, this wavelength is outside the highest gain region of the FP laser.  The range of currents over which single-mode SIL is possible is quite small and further increases in the diode current (State III, bottom panel) cause a transition to a multimode regime similar to state I. Importantly, the condition of single-mode SIL does not necessarily correspond to generation of a soliton or even a MI comb state.  When microcombs were generated, they were typically MI microcombs, generated with a multi-mode FP state; see Fig. \ref{fig:Fig4}(c).

\section{Conclusion}

We have presented hybrid InP and SiN integration of an octave-spanning frequency comb in support of future precision-timing applications.  Our system consists of a monolithic DFB laser butt-coupled to a SiN photonic chip, containing a GVD-engineered Kerr microresonator.  SIL soliton microcombs use back-reflected Rayleigh scattering to modify the pump laser behavior in favor of soliton stability. This process passively stabilizes detuning against thermal instabilities. The advantages of this system include integration, lower pump power requirements to obtain certain soliton spectral profiles, and a simplified soliton experimental protocol, which manifests by a dynamic attraction of detuning to the range for stable soliton propagation. However, the disadvantages of a fundamental reduction in the range of accessible detuning, which reduces the intensity of DWs, and reduced capability for frequency control of the soliton microcomb, which is essential for optical-frequency metrology \cite{drake2019terahertz}. For small Rayleigh scattering rates, the detuning is significantly restricted compared to the maximum value predicted by the LLE.  We reveal a coupling between the linear scattering rate and detuning that allows access to larger detunings. Since the goal of an integrated microcomb for precision metrology would affect numerous scientific and technical research directions, we have analysed potential system strategies and compared their expected performance, using numerical LLE simulations. Agile laser frequency control of a current-modulated, optically isolated DFB laser enables full detuning control to access intensive DW formation, but this approach likely requires more challenging photonic integration. Our modeling predicts that spontaneous soliton generation in PhCR resonators is capable of generating intense DW spectra, potentially with ultralow-noise hybrid integrated lasers\cite{stern2020ultra,huang2019high,morton2018high}. Our results are critical for the future development of chip-scale soliton frequency combs that can be realistically self-referenced.

\section{Funding Information}
This project was funded by the DARPA ACES and DODOS programs and NIST. 

We thank Freedom Photonics for providing the DFB laser chip, Jizhao Zhang and Grisha Spektor for valuable comments on the manuscript, and Andrey S. Voloshin, Igor A. Bilenko and Nikita M. Kondratiev for helpful discussions.

\section{Data Availability}

The data that support the findings of this study are available from the corresponding author upon reasonable request.



%
%

%


\bibliography{brilesBIBv24}

\begin{thebibliography}{44}
\bibitem [{NIS()}]{NISTcommercialSTATEMENT}%
  \BibitemOpen
  \href@noop {} {\enquote {\bibinfo {title} {Mention of specific companies or
  trade names is for scientific communica-tion only, and does not constitute an
  endorsement by nist.}}\ }\BibitemShut {NoStop}%
\end{thebibliography}

\end{document}